\documentclass[aps,pra,amsmath,amssymb,11pt,final,tightenlines,twoside,onecolumn,nofloats,nofootinbib,superscriptaddress,
showkeys,showkeywords]{revtex4-2}

\usepackage[T2A]{fontenc}
\usepackage[utf8x]{inputenc}
\usepackage[russian,english]{babel}
\usepackage{graphicx}% Include figure files
\usepackage{multirow}

\input{journals.def}

\usepackage{xcolor} %для пометок красным
\usepackage{array} %для того, чтоб картинки размещать в одну строку
\usepackage{hyperref} %чтоб ссылки вставить

\begin{document}
	
	\selectlanguage{english}
	
	\noindent {\it ASTRONOMY REPORTS, 2025, Vol. , No. }
	\bigskip\bigskip  \hrule\smallskip\hrule
	\vspace{35mm}
	
	
	\title{The Likelihood of Hosting Undetected Brown Dwarfs in the Solar Vicinity}
	
	\author{\bf\copyright\ 2025 \firstname{D.~R.}~\surname{Bakirova}}
	\email{dianabak@inbox.ru}
	\affiliation{Moscow Institute of Physics and Technology, 141700 Moscow Region,
		Dolgoprudny, Institutskiy Pereulok, 9}
	\affiliation{Institute of Astronomy, Russian Academy of Sciences, Moscow, Russia}
	
	\author{\bf\firstname{O.~Yu.}~\surname{Malkov}}
%	\email{malkov@inasan.ru \textcolor{red}{(я решила не убирать вашу почту, но если это ни к чему, то уберите :))}}
	\affiliation{Institute of Astronomy, Russian Academy of Sciences, Moscow, Russia}
	
	\begin{abstract}
		\vspace{3mm}
		\received{31.12.25}
		\revised{31.12.25}
		\accepted{31.12.25} 
		\vspace{3mm}
		
		Abstract
		\\
Based on the spatial distribution of objects in the solar neighborhood with a radius of 20 parsecs,
and after correcting for the incompleteness of observational data, an expression was obtained for
estimating the probability of finding an object at a given distance from the Sun.
According to these estimates, with a probability of about 0.5, there exists a brown dwarf
in the immediate solar vicinity (< 1.2 pc).
The possible multiplicity of this hypothetical object is discussed,
as well as the reasons why it has not yet been detected.
%		\\
%		Общие замечания по структуре всего этого бардака. \textcolor{red}{Красный текст} -- вопросы к Олегу Юрьевичу. \textcolor{teal}{Сине-зеленый текст} -- вопросы ко мне, которые я отложила на потом. \textit{Курсивом} написано краткое содержание текста, который предполагается в этом месте. 
		
	\end{abstract}
	
	\maketitle
	
\section{INTRODUCTION}

Objects in the solar neighborhood are of great interest to science,
as their proximity allows us to obtain more complete and promising data about the Universe.
According to the currently accepted initial mass function \cite{2024arXiv241007311K},
the overwhelming majority of stars in the Galaxy are red and brown dwarfs.
In this context, it is noteworthy that at present the nearest known object to the Solar System
is the triple system $\alpha$ Cen, consisting of one red dwarf and two yellow dwarfs.
Either this is related to peculiarities of star formation near the Sun, or observational selection effects are at play.
% согласно статистике, между системой $\alpha$ Cen и Солнцем находится еще не обнаруженный нами коричневый карлик?

This question is investigated in the present work.
In Section \ref{sec:data}
% части II мы, используя каталог Kirkpatrick et al \textcolor{red}{(тут можно не приводить ссылку?)},
distances between nearest neighbors are calculated for objects in the solar neighborhood (up to 20 pc)
and from the resulting distribution, a relationship is derived between distance and the probability of the existence of an object
at that distance and closer.
In Section \ref{sec:probability}, the obtained results are used to estimate the probability
of the existence of an object at a distance from the Sun smaller than the distance to the system $\alpha$ Cen (\ref{sec:distance}),
as well as to estimate the multiplicity of this object (\ref{sec:multiplicity}).
In Section \ref{sec:reasons}, possible reasons are discussed
why an object at such a close distance still remains undetected.
Conclusions are presented in Section \ref{sec:conclusions}.

%\section{ОЦЕНКА ВЕРОЯТНОСТИ СУЩЕСТВОВАНИЕ ЗВЕЗДЫ НА ЗАДАННОМ РАССТОЯНИИ ОТ СОЛНЦА И БЛИЖЕ}
\section{MEAN DISTANCE BETWEEN STARS IN THE SOLAR NEIGHBORHOOD}
\label{sec:data}

\subsection{SOURCE DATA}

At present, there exist in the literature several sources of information (catalogs) about nearby and/or
cool stars, two of which are of particular interest due to their representativeness and
variety of presented data.

First and foremost is the catalog of Kirkpatrick et al.
\footnote{\url{https://vizier.cds.unistra.fr/viz-bin/VizieR?-source=J/ApJS/271/55}}
\cite{2024ApJS..271...55K}, which includes objects located at distances up to 20 pc from the Sun.
% В нашей работе используются данные из каталога Kirkpatrick et al
% (\url{https://vizier.cds.unistra.fr/viz-bin/VizieR?-source=J/ApJS/271/55}),

In addition, there is the catalog of Best et al.
\footnote{\url{https://zenodo.org/records/10573247}}
%Best et al (\url{https://docs.google.com/spreadsheets/d/1i98ft8g5mzPp2DNno0kcz4B9nzMxdpyz5UquAVhz-U8/edit?gid=1795845968#gid=1795845968}),
\cite{2021AJ....161...42B}.
This catalog is called UltracoolSheet and contains data
(positions, proper motions, parallaxes, photometry, multiplicity,
spectroscopic classification, etc.) on more than 4000 ultracool dwarfs
of spectral types M6 and later.

\begin{figure}[h]
  \includegraphics[width=0.45\textwidth]{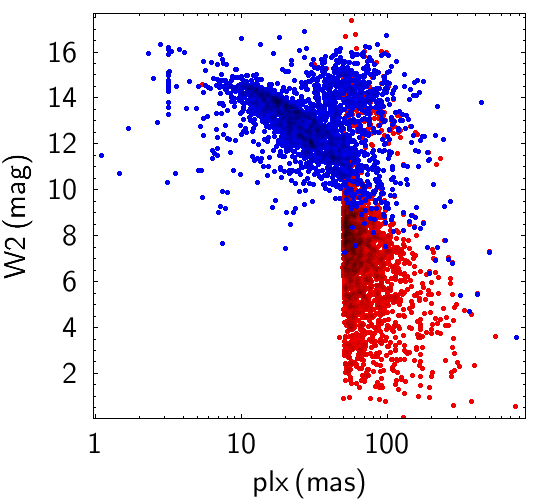}
  \includegraphics[width=0.45\textwidth]{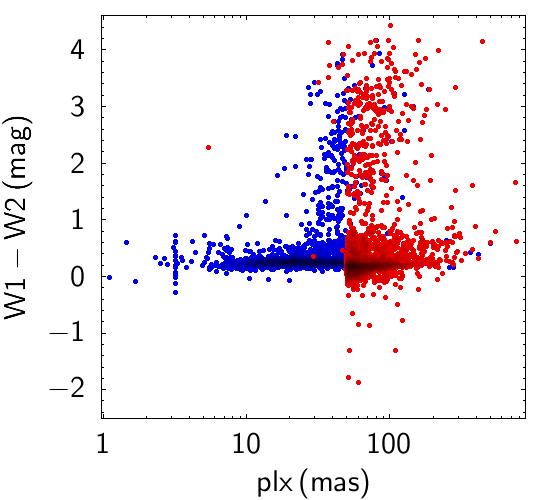}
  \caption{$W2$ magnitudes (left) and $W1-W2$ color indices (right) as a function of parallax
for objects from the catalog of Kirkpatrick et al. \cite{2024ApJS..271...55K} (red)
and the catalog of Best et al. \cite{2021AJ....161...42B} (blue).}
% По осям ординат отложены звездная величина в фильтре W1 (а) и разность звездных величин в фильтрах W1 и W2 (б). В следствие того, как холодные объекты расположена на диаграмме Герцшпрунга-Рассела, можно говорить о почти линейной зависимости этих показателей от светимости и температуры объектов соответственно. Следовательно, графики показывают, что каталог Kirkpatrick et al полон в определенном радиусе, в отличие от каталога Best et al.}
\label{fig:kirkbest}
\end{figure}

Figure \ref{fig:kirkbest} presents the distribution of objects from these two catalogs
by parallax, $W2$ magnitude, and $W1-W2$ color index.
Obviously, for solving our problem, the first of these sources, Kirkpatrick et al. \cite{2024ApJS..271...55K}, is more suitable,
since it better ensures spatial completeness of objects.
We use this catalog in what follows.

%\begin{figure*}[h]
%	\begin{tabular}{>{\centering}p{0.5\textwidth}>{\centering\arraybackslash}p{0.5\textwidth}}
%		а)
%		\includegraphics[width=0.3\textwidth]{W2.png}
%		
%		&
%		
%		б)
%		\includegraphics[width=0.3\textwidth]{W1-W2.png}
%	\end{tabular}
%	
%	\caption{На графиках изображены объекты каталогов Kirkpatrick et al (красные) и Best et al (синие). По осям ординат отложены звездная величина в фильтре W1 (а) и разность звездных величин в фильтрах W1 и W2 (б). В следствие того, как холодные объекты расположена на диаграмме Герцшпрунга-Рассела, можно говорить о почти линейной зависимости этих показателей от светимости и температуры объектов соответственно. Следовательно, графики показывают, что каталог Kirkpatrick et al полон в определенном радиусе, в отличие от каталога Best et al.}
%\end{figure*}

The catalog of Kirkpatrick et al. \cite{2024ApJS..271...55K} contains information about 4407 objects.
Excluded from consideration were 17 objects located farther than 20 pc from the Sun,
as well as 72 objects for which parallaxes are unknown.
The remaining 4318 objects were combined into 2672 systems, where a system is
a grouping of one or more stars gravitationally bound to each other.
Table \ref{tab:multiplicity} shows the percentage distribution of systems depending on their multiplicity.

\begin{table}[h]
	\centering
	\caption{Percentage distribution of systems in \cite{2024ApJS..271...55K} depending on their multiplicity.}
%	\label{tab:my-table}
	\begin{tabular}{|c|c|c|}
		\hline
		Multiplicity & Number & Fraction                   \\ \hline
		1            & 2001   & 75\%                       \\ \hline
		2            & 543    & 20\%                       \\ \hline
		3            & 110    & 4\%                        \\ \hline
		4            & 12     & \multirow{3}{*}{$\sim$1\%} \\ \cline{1-2}
		5            & 5      &                            \\ \cline{1-2}
		6            & 1      &                            \\ \hline
	\end{tabular}
\label{tab:multiplicity}
\end{table}

\begin{figure*}[h]
  \includegraphics[width=0.45\textwidth]{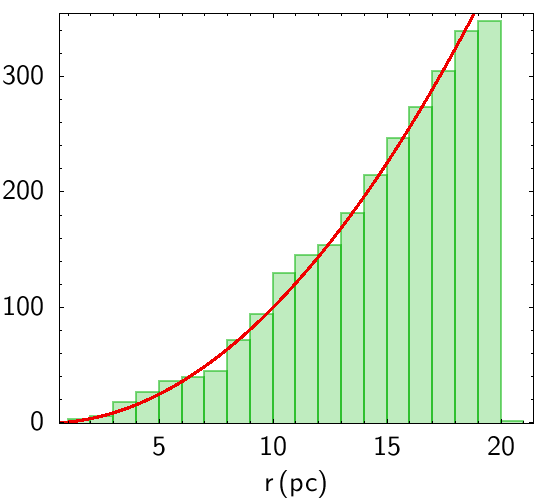}
  \includegraphics[width=0.45\textwidth]{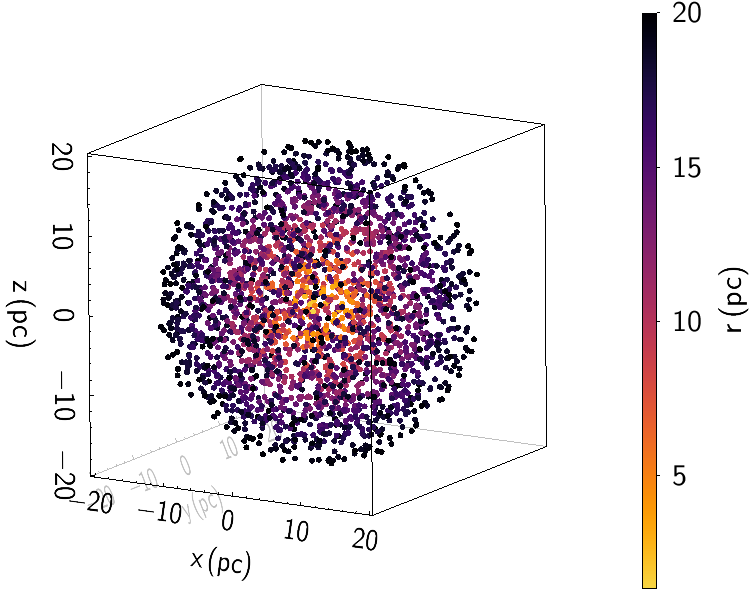}
%	\begin{tabular}{>{\centering}p{0.5\textwidth}>{\centering\arraybackslash}p{0.5\textwidth}}
%		а)
%		\includegraphics[width=0.3\textwidth]{hist_r.png}% Вот как импортировать EPS art
%		&
%		б)
%		\includegraphics[width=0.4\textwidth]{3d.png}
%	\end{tabular}
	\caption{Distribution of systems by their distance from the Sun (left), the red curve shows the theoretical dependence $N \varpropto r^2$. Spatial distribution of systems relative to the Sun (right).}
\label{fig:space}
\end{figure*}

The spatial distribution of systems is shown in Fig. \ref{fig:space}.
Obviously, one can speak
% Рис. 2 наглядно показывает, что мы можем говорить 
of a uniform distribution of systems in the studied neighborhood.

\subsection{ESTIMATION OF THE MEAN DISTANCE TO THE NEAREST NEIGHBOR}

\begin{figure}[h]
  \includegraphics[width=0.45\textwidth]{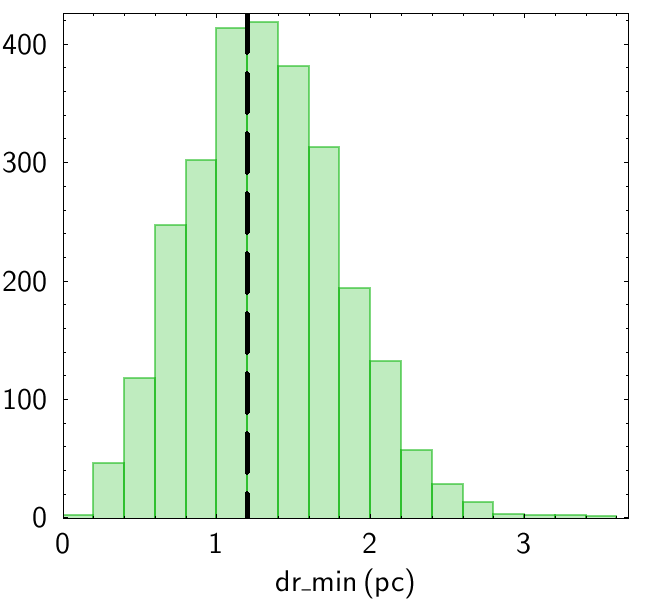}
  \includegraphics[width=0.45\textwidth]{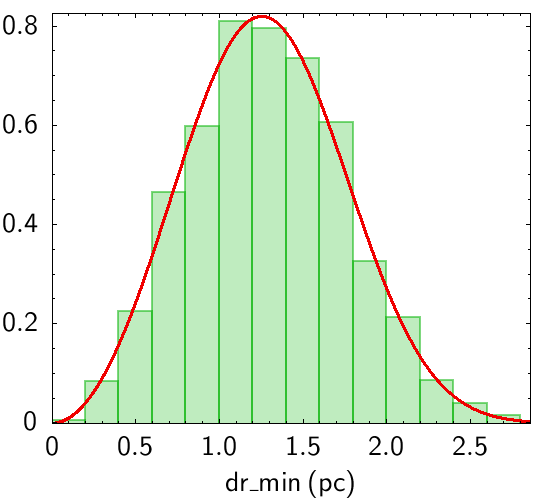}
%	\center{\includegraphics[width=0.8\textwidth]{hist_dr_min.png}}
	\caption{Distribution of systems by their distances to the nearest neighbor.
Left: the peak of the distribution is marked with a black dashed line.
Right: corrected for incompleteness, red curve shows the Hertz dependence \eqref{equ:hertz}.}
\label{fig:drmin}
\end{figure}

For each system, the distance to the nearest neighbor (dr\_min) was found, the resulting distribution
is shown in Fig. \ref{fig:drmin}. The peak of the distribution (corresponding to approximately 1.2 pc)
can be taken as the mean distance to the nearest neighbor.

%\begin{figure}[h]
%	\center{\includegraphics[width=0.3\textwidth]{hist_min_dr_full_gerz.png}}
%	\caption{Распределение систем по их расстоянием до ближайшего соседа.
% Красная кривая -- зависимость Гертца \eqref{equ:hertz}.}
%\label{fig:hertz}
%\end{figure}

\begin{figure*}[h]
  \includegraphics[width=0.45\textwidth]{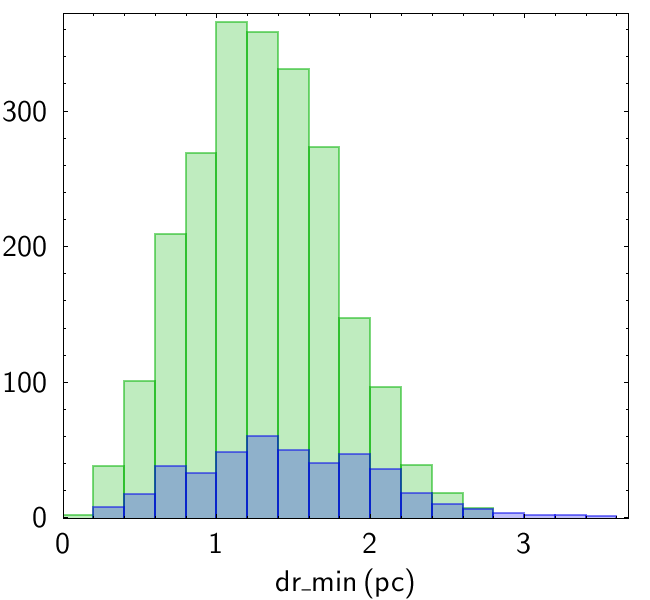}
  \includegraphics[width=0.45\textwidth]{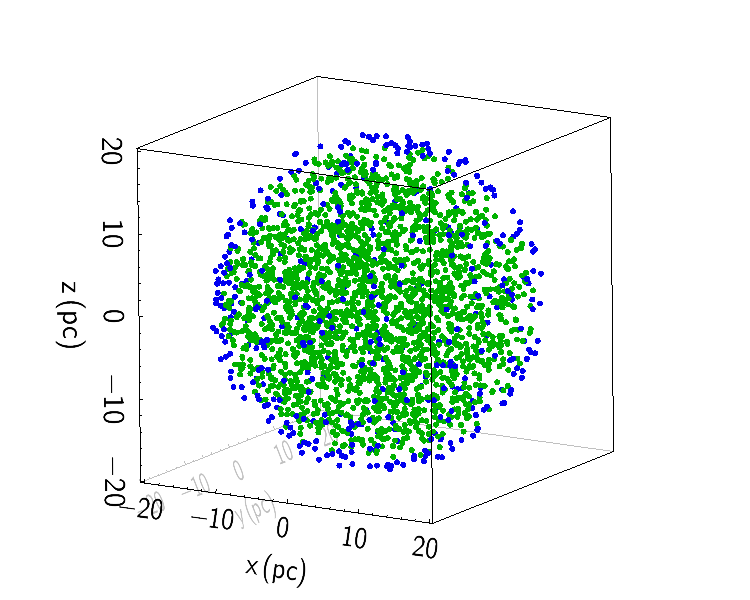}
%	\begin{tabular}{>{\centering}p{0.5\textwidth}>{\centering\arraybackslash}p{0.5\textwidth}}
%		а)
%		\includegraphics[width=0.3\textwidth]{hist_min_dr_агдд.png}% Вот как импортировать EPS art
%		&
%		б)
%		\includegraphics[width=0.4\textwidth]{3d_full.png} 
%	\end{tabular}
	\caption{Correction for incompleteness: exclusion from consideration of systems
located at the boundary of the studied region, in a shell of width 1.2 pc.
They are marked in blue on the distribution of systems by their distances to the nearest neighbor (left)
and on the spatial distribution of systems relative to the Sun (right).}
\label{fig:shell}
\end{figure*}

It is advisable to account for the incompleteness of our data, for which it is necessary to exclude systems
located at the boundary of the studied region, namely in a shell of width
equal to the mean distance to the nearest neighbor, i.e., 1.2 pc (see Fig. \ref{fig:shell}).

To verify our calculations and the completeness of the data, we compared the resulting distribution
with the nearest neighbor problem, which was studied by Hertz \cite{1943RvMP...15....1C}.
According to his result,
% В его работе данная задача была рассмотрена со стороны статистики, и полученный результат гласил, что
the dependence of the probability of finding the nearest neighbor at a given distance on the distance itself
is described by the expression:

\begin{equation}
\omega(r) = \text{exp}(-4\pi r^3 n/3) \cdot 4\pi r^2 n,
\label{equ:hertz}
\end{equation}
%\[\omega(r) = \text{exp}(-4\pi r^3 n/3) \cdot 4\pi r^2 n\]
where $n$ is the number of objects per unit volume.
Having, after correcting for incompleteness, 2253 systems in a sphere of radius 18.8 pc,
we can construct this dependence. After normalizing our histogram by area, it can be seen
that our data agree well with the theoretical estimate of Hertz (see Fig. \ref{fig:drmin}).
His work also provided the mean distance to the nearest neighbor:

% n = 0.08098

\begin{equation}
\overline{dr\_min} = 0.55396 n^{-1/3}
\end{equation}
%\[\overline{dr\_min} = 0.55396n^{-1/3}\]
Substituting our density $n$, we obtain $\overline{dr\_min} \simeq 1.28$ pc, which is similar to our estimate.

% шрифт 28 в topcat

%При этом существующая на данный момент оценка среднего расстояния между звездами в ближайшей солнечной окрестности равна примерно 1.9 пк \cite{1978SvA....22..186L}, что больше расстояния до ближайшего к Солнцу соседа (система $\alpha$ Cen, $\sim 1.3$ пк \cite{2016A&A...586A..90P}) и больше, чем наша и теоретическая оценки среднего расстояния до ближайшего соседа. \textcolor{red}{Статья про среднее расстояние между звездами от 1978 года, я понимаю, что это абсурд :) Но у меня не получилось найти новее, к сожалению. Плохо у меня пока получается искать статьи. И еще. Я не очень понимаю, как это по смыслу еще сюда приплести и какой вывод сделать. Ну вот просто интересный факт и все...}

%P.Hertz, Math. Ann. 67, 387 (1909)

\begin{figure}[h]
  \includegraphics[width=0.45\textwidth]{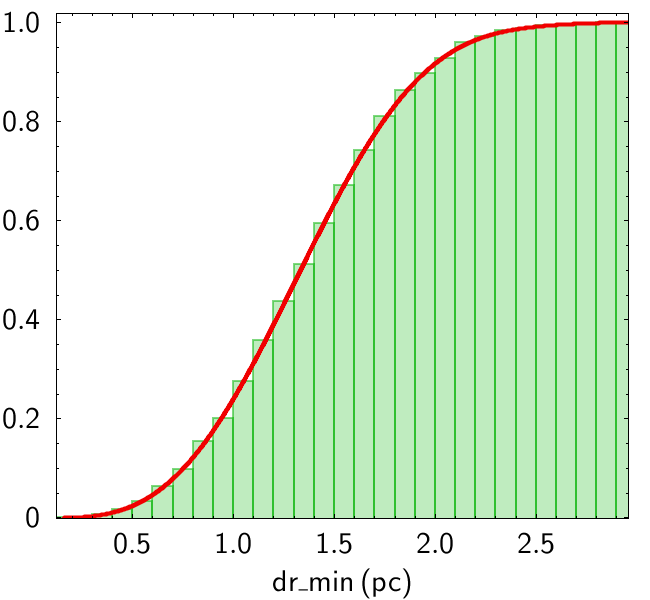}
  \includegraphics[width=0.45\textwidth]{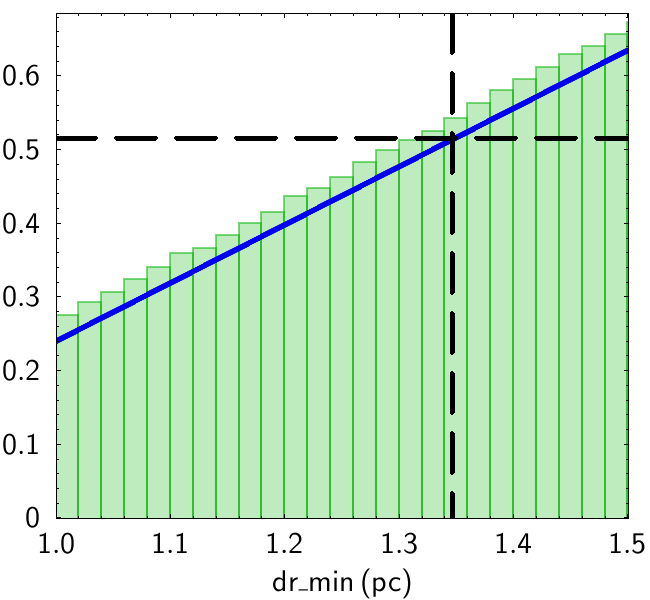}
%	\center{\includegraphics[width=0.45\textwidth]{hist_dr_min_approx.png}}
	\caption{Left: cumulative distribution of systems by their distances to the nearest neighbor,
normalized to unity.
Red curve shows the approximating polynomial \eqref{equ:integral}.
Right: fragment of the distribution \ref{fig:integral} in the range 1 -- 1.5 pc.
Blue line shows the linearized approximating polynomial \eqref{equ:line} on this segment,
black dashed line shows the position of the system $\alpha$ Cen.}
\label{fig:integral}
\end{figure}

Figure \ref{fig:integral} presents the {\it cumulative}
distribution of distances to the nearest neighbor, normalized to unity,
i.e., the probability of the existence of an object located at the indicated distance from the Sun and closer.
Using the least squares method, it can be approximated by a polynomial:

\begin{eqnarray}
	P(x) = -1.6\cdot10^{-6} - 6.9\cdot10^{-3}x + 4.3\cdot10^{-2}x^2 - 1.1\cdot10^{-1}x^3 + 7\cdot10^{-1}x^4 \nonumber \\ - 3.8\cdot10^{-1}x^5 - 1.1\cdot10^{-1}x^6 + 1.3\cdot10^{-1}x^7 -  3.7\cdot10^{-2}x^8 + 3.4\cdot10^{-3}x^9
\label{equ:integral}
\end{eqnarray}

%\begin{figure}[h]
%	\center{\includegraphics[width=0.8\textwidth]{hist_dr_min_approx_lin.png}}
%	\caption{Фрагмент распределения \ref{fig:integral} на участке 1 -- 1.5 пк.
%%Интегральное распределение систем по их расстояниям до ближайшего соседа, нормированное на единицу,
%% на отрезке от 1 пк до 1.5 пк.
% Синяя линия -- линеаризованный на этом отрезке аппроксимирующий многочлен \eqref{equ:line},
% черным пунктиром показано положение системы $\alpha$ Cen. }
% \label{fig:line}
%\end{figure}

\subsection{ERROR ESTIMATION}

\begin{figure}[h]
	\center{\includegraphics[width=0.45\textwidth]{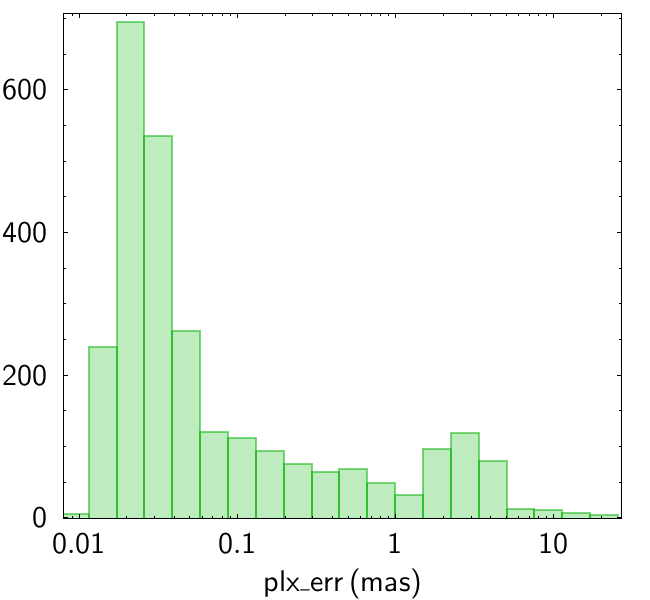}}
	\caption{Distribution of parallax uncertainties for objects from the catalog of Kirkpatrick et al. \cite{2024ApJS..271...55K}}
\label{fig:errors}
\end{figure}

To estimate the error of the obtained dependence, one should first of all account for observational errors,
which often have quite complex distributions. Thus, Fig. \ref{fig:errors}
shows the parallax uncertainties.

Another source of uncertainty should also be noted:
%На данный момент можно сказать о том, что наибольший вклад в ошибку вносит 
the incompleteness of the catalog of Kirkpatrick et al. \cite{2024ApJS..271...55K} with respect to faint stars.
However, possible completion of the catalog will shift the obtained distribution (Fig. \ref{fig:integral})
only to the left, which will lead only to an increase in the probability corresponding to each value of distance
to the nearest neighbor. This means that the value obtained in this work
represents a lower estimate, which is also a very significant result.

\section{APPLICATION OF THE OBTAINED RESULTS TO THE NEAREST SOLAR VICINITY}
\label{sec:probability}

\subsection{PROBABILITY OF EXISTENCE OF A NEW NEAREST OBJECT}
\label{sec:distance}

Linearization of the polynomial \eqref{equ:integral} in the range from 1 pc to 1.5 pc (see Fig. \ref{fig:integral}),
simplifies it to

\begin{eqnarray}
	f(x) = 0.79x - 0.55
\label{equ:line}
\end{eqnarray}

One can see that the probability of the existence of an unknown object at a distance
up to the system $\alpha$ Cen $r \simeq 1.346$ pc \cite{2016A&A...586A..90P} and closer
is equal to

% Подставив расстояние до системы $\alpha$ Cen $r \simeq 1.346$ пк \cite{2016A&A...586A..90P}, мы получили вероятность существования объекта на этом расстоянии и ближе, равную:

\[(51.5 \pm 0.2)\% \]

The probability uncertainty is estimated from the parallax measurement uncertainties.

\subsection{POSSIBLE MULTIPLICITY OF THE NEW OBJECT}
\label{sec:multiplicity}

\begin{figure}[h]
	\center{\includegraphics[width=0.45\textwidth]{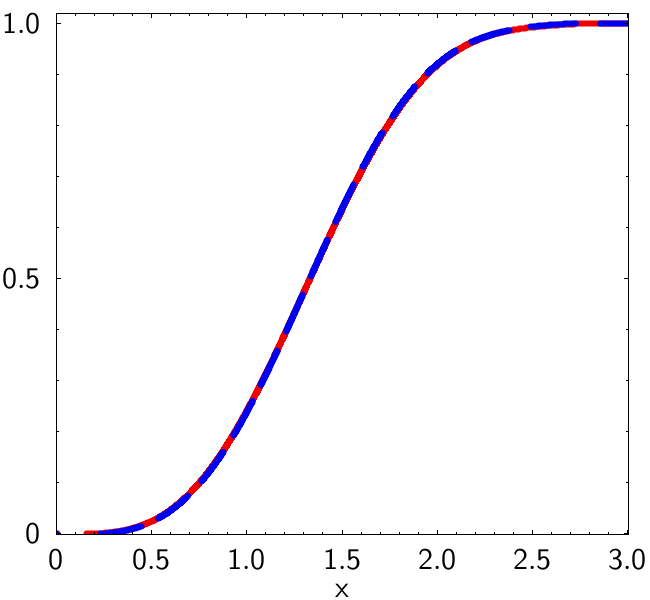}}
	\caption{Approximating polynomials for the cumulative distributions of single (red) and multiple (blue) systems by their distances to the nearest neighbor, normalized to unity.}
\label{fig:multiple}
\end{figure}

% Также все 
The calculations described above were performed separately for single and multiple systems.
As can be seen from Fig. \ref{fig:multiple}, the obtained polynomials approximating the cumulative distribution
coincide with each other with very high accuracy, which indicates that single and multiple systems are distributed
in a similar manner. In accordance with Table \ref{tab:multiplicity}, this means that the probability
that the new object will be multiple is equal to $\sim25\%$.

%\subsection{ПОЧЕМУ ДО СИХ ПОР НЕ ОБНАРУЖЕН? ВОЗМОЖНЫЕ ПРИЧИНЫ}
\subsection{ON THE REASONS WHY THE NEW OBJECT HAS NOT BEEN DETECTED}
\label{sec:reasons}

It is logical to assume that if such a close object to the Sun has not yet been detected,
it must be faint and/or cool. Therefore, it can be assumed that we are dealing with a brown dwarf,
which corresponds to the considerations presented in the Introduction.
%также сходится с нашими размышлениями в введении -- странно, что ближайшим к Солнцу объектом является система из одного красного и двух желтых карликов, хотя коричневых карликов в нашей Вселенной намного больше. 

At present, the nearest brown dwarf to us is a relatively bright object -- WISE J104915.57-531906.1 (known as Luhman 16), a binary system of two brown dwarfs of spectral classes L8$\pm$1 and T1$\pm$2, located at a distance of approximately 2 pc \cite{2024AN....34530158B}.
% \textcolor{red}{В этой статье астрометрия была получена по обзорам Хаббла, если я верно поняла. Это тогда плохой пример в контексте того, что я писала выше?}.

Brown dwarfs are relatively bright objects in the infrared range, therefore,
speaking of photometric data confirming the spectral classes of brown dwarfs,
it is advisable to refer to the WISE survey \cite{2010AJ....140.1868W}.
Table \ref{tab:BD} presents absolute magnitudes in the W1 (WISE) and G (Gaia) bands,
corresponding to brown dwarfs. Speaking of the sought object, we can assume
that it might not have been included in the WISE survey due to the fact that this survey has few time series
for different regions of the sky, and the object could be too faint and not be registered.
But even if it is present in the WISE survey, its photometric data are insufficient for unambiguous classification,
since the photometry of brown dwarfs is very similar to, for example, the photometry of Mira variables.
% \cite{2024PZ.....44...54A}.
% \textcolor{red}{Наверное тут нужна ссылка на подтверждение этой мысли} 

It is necessary to take into account the circumstance
that if the sought object is a brown dwarf, it will most likely be too faint
for the main source of astrometric data -- the Gaia DR3 survey, whose limiting magnitude
for determining parallaxes corresponds to approximately $20^{\text{m}}$ \cite{2023ARep...67..288M}.
As can be seen from Table \ref{tab:BD}, for spectral classes T and Y the absolute magnitudes in the G band
are sufficiently large, i.e., objects of spectral class Y will not be detected by Gaia
even being in the Oort cloud \cite{2023ARep...67..288M}.
All this means that the sought object may be present in modern photometric surveys,
but its parallax and proper motion remain unknown
% искомого объекта, поэтому мы и не можем сказать о его близости к Солнцу.

\begin{table}[]
	\centering
	\caption{Estimates of absolute magnitudes of cool stars in the filters of Gaia and WISE surveys (from \cite{2023ARep...67..288M}).}
% 	\label{tab:my-table}
	\begin{tabular}{|c|c|c|}
		\hline
		Spectral class & G, Gaia & W1, WISE \\ \hline
		L5                 & 18.50   & 11.19    \\ \hline
		T4.5               & 21.04   & 12.97    \\ \hline
		Y2                 & 32.2    & 18.1     \\ \hline
	\end{tabular}
\label{tab:BD}
\end{table}

\section{CONCLUSIONS}
\label{sec:conclusions}

Based on data from the catalog of objects in the nearest (up to 20 pc) solar neighborhood
of Kirkpatrick et al. \cite{2024ApJS..271...55K}, a dependence of the probability of the existence of an object
at a given distance from the Sun and closer on the distance itself was obtained, see \eqref{equ:integral}.
After linearization of \eqref{equ:integral} in the range from 1 pc to 1.5 pc, from the obtained dependence \eqref{equ:line}
% было подставлено расстояние до ближайшего к Солнцу соседа, системы $\alpha$ Cen, из чего
the probability of the existence of an object located between the system $\alpha$ Cen
and the Sun was estimated: 51.5 $\pm$ 0.2\%.
The probability that this hypothetical object is a binary/multiple system is approximately 25\%.
%Мы заметили подобность распределений одиночных и кратных систем в ближайшей солнечной окрестности, что позволяет сказать о вероятности того, что новый объект будет кратной системой, соответствующей примерно 25\%, то есть доле кратных систем. 

The paper also discusses the reasons why
% Также были приведены рассуждения, почему
such a close object may not yet be detected: assuming it should be a brown dwarf,
it can be expected that it either turned out to be too faint to be registered
(an object of spectral class Y), or it is present in infrared surveys (such as WISE), but is too faint
for obtaining its astrometric parameters by the Gaia observatory.

Plans include refining the error value of the obtained results, taking into account uncertainties of the used data,
as well as repeating the calculations for the brown dwarf catalog currently being developed
for the Euclid survey \cite{2025arXiv250322497V} with the aim of verifying the results obtained here.

\section*{ACKNOWLEDGMENTS}

We thank A. Rastorguev and M. Prokhorov
%его замечания, а также \textcolor{red}{Расторгуева и Нуриева (как разобрала в ваших заметках с вопросами)}
for valuable comments, thanks to which the article was significantly improved.

\section*{Conflict of Interests}

The authors declare that they have no conflicts of interest.

\clearpage

\bibliographystyle{maik}
\bibliography{br_dw}

\clearpage

\selectlanguage{english}
\begin{center}\bfseries The likelihood of hosting undetected brown dwarfs in solar vicinity
\end{center}
\begin{center}
{\bf Bakirova D. R.$^1, 2$, Malkov O. Yu.$^2$}

$^1$ Moscow Institute of Physics and Technology, Moscow 117303, Russia

$^2$ Institute of Astronomy of the RAS, Moscow 119017, Russia\\[1cm]

\begin{minipage}{\textwidth}
\small 
Abstract.
Based on the spatial distribution of objects in the solar neighborhood with a radius of 20 pc,
 and after correcting for incomplete observational data, an expression was derived
 for estimating the probability of finding an object at a given distance from the Sun.
According to these estimates, there is a probability of approximately 0.5 that a brown dwarf exists
 in the immediate solar neighborhood ($<$1.2 pc).
The possible multiplicity of this hypothetical object is discussed,
 as well as the reasons why it has not yet been detected.
% \textcolor{red}{(I thought that since you're writing an abstract for the Russian part, then you're probably writing here too. It's just a translated abstract, right?)}
\end{minipage}
\end{center}

\begin{center}
\begin{minipage}{\textwidth}

Keywords: {\it brown dwarf, solar neighbourhood}\\

\end{minipage}
\end{center}

\end{document}